\documentclass{arxiv}
\pdfoutput=1	
\copyrightyear{2005}
\pubyear{2005}
\application

\usepackage{multirow}
\usepackage{booktabs}
\usepackage{array}
\usepackage{siunitx}
\usepackage{placeins}
\usepackage[ruled]{algorithm2e}

\begin{document}
\firstpage{1}

\title[Real-time multi-view deconvolution]{Real-time multi-view deconvolution}
\author[Schmid \textit{et~al}]{Benjamin Schmid\,$^{1,*}$ and Jan Huisken\,$^{1,}$\footnote{to whom correspondence should be addressed}}
\address{$^{1}$Max Planck Institute of Molecular Cell Biology and Genetics, 01307 Dresden, Germany}

\history{Received on XXXXX; revised on XXXXX; accepted on XXXXX}

\editor{Associate Editor: XXXXXXX}

\maketitle

\begin{abstract}

\section{Summary:}
In light-sheet microscopy, overall image content and resolution are improved by acquiring and fusing multiple views of the sample from different directions. State-of-the-art multi-view (MV) deconvolution employs the point spread functions (PSF) of the different views to simultaneously fuse and deconvolve the images in 3D, but processing takes a multiple of the acquisition time and constitutes the bottleneck in the imaging pipeline. Here we show that MV deconvolution in 3D can finally be achieved in real-time by reslicing the acquired data and processing cross-sectional planes individually on the massively parallel architecture of a graphics processing unit (GPU).
\section{Availability:}
Source code and binaries are available on github (\href{https://github.com/bene51/}{https://github.com/bene51/}), native code under the repository 'gpu\_deconvolution', Java wrappers implementing Fiji plugins under 'SPIM\_Reconstruction\_Cuda'. 

\section{Contact:} \href{bschmid@mpi-cbg.de}{bschmid@mpi-cbg.de}

\section{Supplementary information:} Supplementary data are available at the end of this document.
\end{abstract}

\section{Introduction}

MV imaging is particularly useful in light-sheet microscopy where consecutive views are acquired in short succession, allowing reconstruction of entire developing organisms without artifacts~\citep{Huisken04}. Due to the low photo-toxicity in light sheet microscopy, time-lapse experiments are oftentimes run over days and TBs of data accumulate quickly. MV fusion is therefore particularly desirable to be performed in real-time to eliminate redundant information from different views. Best fusion results, however, are achieved by combining fusion with 3D deconvolution~\citep{Swoger07,Verveer07,Shroff13}. Although efficient Bayesian MV deconvolution based on the Richardson-Lucy (RL) algorithm has been shown recently to outperform existing methods in terms of fusion quality and convergence speed, it is still too slow for real-time processing of typical data volumes~\citep{Preibisch14}. 

The RL deconvolution iterations consist only of convolutions and pixel-wise arithmetic operations and could therefore be significantly accelerated using dedicated hardware such as a GPU. The large memory requirements of MV deconvolution, however, exceed the limited resources of modern GPUs even for moderate data sizes (Supplementary Note 1). Previous attempts therefore required splitting the data into blocks of appropriate size. Each block then either had to be transferred to and from the GPU in each RL iteration~\citep{Preibisch14}, or blocks needed to share a considerable amount of overlap to avoid border artifacts~\citep{Ott11}. Therefore, GPU-based implementations only achieved a three-times performance gain~\citep{Preibisch14}.

\begin{figure*}
\centerline{\includegraphics{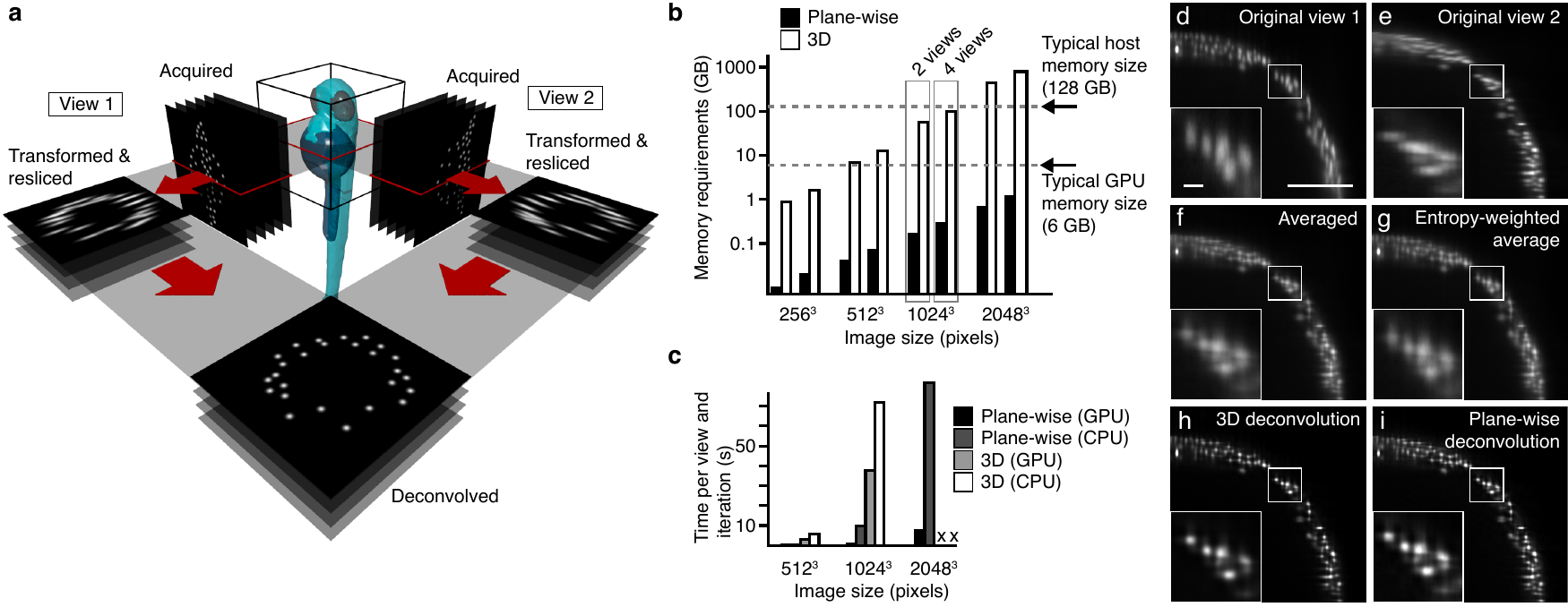}}
\caption{Plane-wise multi-view deconvolution concept and performance.
\textbf{(a)} Concept of plane-wise deconvolution for two views. Each data set is resliced into planes orthogonal to the microscope's rotation axis. Data sets are deconvolved plane-by-plane. \textbf{(b)} Memory requirements for traditional 3D and our plane-wise multi-view deconvolution, for various data sizes and numbers of views, on a logarithmic scale. \textbf{(c)} Execution times for plane-wise multi-view deconvolution, implemented on GPU and CPU, and 3D deconvolution, with and without GPU support. Memory requirements for 3D deconvolution timings for the $2048^3$ pixel data set were beyond the capabilities of our workstation. \textbf{(d-i)} Resulting images of a 9 hours post fertilization transgenic \textit{Tg(h2afva:h2afva-mCherry)} zebrafish embryo, using different methods (view along the rotational axis, scale bar \SI{100}{\micro\meter}, \SI{10}{\micro\meter} in the inset): \textbf{(d, e)} acquired raw data, \textbf{(f-i)} fusion performed by \textbf{(f)} averaging, \textbf{(g)} entropy-weighted averaging, \textbf{(h)} 3D multi-view deconvolution and \textbf{(i)} plane-wise multi-view deconvolution (10 iterations). (Dell T6100, Intel E5-2630 @2.3 GHz 2 processors, 64 GB RAM; Graphics card: Nvidia GeForce GTX TITAN Black).}
\label{fig:01}
\end{figure*}

\section{Results}

The primary goal of MV fusion is the improvement of the poor axial resolution in a single 3D dataset using the superior lateral resolution of an additional, overlapping dataset, and not necessarily to improve resolution beyond the intrinsic lateral resolution. We therefore approximated the full 3D transfer function with a 2D PSF, neglecting one lateral component (along the rotation axis), and processed each plane orthogonal to the rotation axis independently (Fig.~1a). Memory requirements were thereby reduced by the number of lines read out from the camera chip, i.e. typically 100-1000 fold (Fig.~1b). This allowed us to implement the entire MV deconvolution on a GPU. Taking advantage of three CUDA (Compute Unified Device Architecture) streams, we interleaved GPU computations with data transfers, such that not only expensive copying to and from GPU memory, but also reading and writing data from and to the hard drive came without additional cost (Supplementary Note 2). Compared to 3D MV deconvolution, with and without GPU support, we thereby reduced processing times by a factor of up to 25 and 75, respectively (Fig.~1c, Supplementary Table 1), while producing comparable results.

We compared the results obtained by our plane-wise deconvolution to the methods commonly used in the light-sheet community, such as established 3D deconvolution~\citep{Preibisch14}, averaging and entropy-based fusion~\citep{Preibisch10}~(Fig.~1d-i). Both averaging and entropy-based fusion were blurry and showed cross-shaped artifacts, originating from the elongated PSFs along the optical axes. 3D deconvolution as well as our plane-wise variant reduced artifacts and enhanced the contrast, thus truly improving the resolution in the fused data set. While our results were comparable to the slow full 3D deconvolution (Fig.~1h,i; Supplementary Fig.~1), processing times and memory requirements were heavily reduced so that the entire deconvolution could be performed in real time.

We provide our software as a C library that can be directly linked to camera acquisition software for real-time processing. To also benefit from the increased performance in post-processing, we additionally created Java wrappers to provide plugins for the open source image processing software Fiji~\citep{Schindelin12} (Supplementary material).

\section{Validation}

Our plane-wise deconvolution approximates 3D deconvolution by neglecting the contribution of the PSF along the rotation axis. Using artificial data (Supplementary Fig.~2), we found that the validity of this approximation is independent of the amount of noise (Supplementary Fig.~3), but depends on the lateral extents of the PSF. Keeping its axial standard deviation fixed at eight pixels, a typical value measured on our microscopes, we found that up to a lateral standard deviation of 2-3 pixels, results from plane-wise and 3D deconvolution are undistinguishable (Supplementary Fig.~4). The measured lateral standard deviation of the PSF was typically between 1.5 and 1.8 pixels on our microscopes.


\section{Conclusion}

With the advent of first commercially available systems, light-sheet microscopy becomes more and more popular. Its photo-efficiency enables long time-lapse imaging of living samples to study fundamental questions in developmental biology. However, the huge data rates and enormous amounts of data it produces also open new challenges for data processing and handling. 
A key problem in light-sheet microscopy is the fusion of data recorded from multiple angles. In this paper, we have presented a new method that performs MV deconvolution plane-wise, which reduces memory requirements compared to existing methods and thus permits an entirely GPU-based implementation. The achieved acceleration makes MV deconvolution for the first time applicable in real-time without the need for data cropping or resampling.

%
%

\section*{Acknowledgement}
We thank all members of the Huisken lab for stimulating discussions.


\paragraph{Conflict of interest\textcolon} none declared.

\bibliographystyle{natbib}
\bibliography{Document}

\onecolumn

\setcounter{equation}{0}
\setcounter{figure}{0}
\setcounter{table}{0}
\setcounter{page}{1}

{\helveticabold\fontsize{16}{21}\selectfont\raggedright Supplementary Material}

\section* {Supplementary Note 1: Memory requirements for multi-view deconvolution}

The multi-view Richardson-Lucy deconvolution algorithm iteratively updates the current estimate of the true image by the following formula:

$$
\psi^{t+1} = \psi^{t}\;\prod_{v \in V}\frac{\phi_v}{\psi^{t} * P_v} \ast P_v^{*}
$$
Assuming the two convolutions are performed in Fourier space, the memory required per pixel (of the isotropic fused dataset) is $22 * v + 12$ bytes, where $v$ is the number of views:\\

\begin{center}
\begin{tabular}  { lcc }    \toprule
                                         &           &  memory requirement \\
description                              & data type &  (bytes/pixel)      \\ \midrule
kernel spectrum FFT$(P_v)$               & complex   &  8 $v$              \\
inverted kernel spectrum FFT$(P_v^{*})$  & complex   &  8 $v$              \\
data $\phi_v$                            & uint16    &  2 $v$              \\
weights                                  & float     &  4 $v$              \\
estimate $\psi$                          & float     &  4                  \\
estimate spectrum FFT$(\psi)$            & float     &  4                  \\
temporary buffer                         & float     &  4                  \\ \midrule
                                         &           &  22 $v$ + 12        \\ \bottomrule
\end{tabular}
\end{center}

\section* {Supplementary Note 2: Implementation}

\subsection* {1. Summary of optimization methods}
\citet{Preibisch14} derived a number of optimizations to make traditional Richardson-Lucy multi-view deconvolution more efficient. While the time required for one iteration remained unchanged, less iterations were required for achieving the same deblurring. The implemented variants all used the following formula, but replaced $X$ with the expressions given below:

$$
\psi^{t+1} = \psi^{t}\;\prod_{v \in V}\frac{\phi_v}{\psi^{t} * P_v} \ast X
$$

\begin{itemize}
   \item Independent:
  \begin{align*}
  X &= P_v^{*}
  \intertext{\item Efficient Bayesian:}
  X &= P_v^{*}\;\prod_{w \in W_v} P_v^{*} \ast P_w \ast P_w^{*}
  \intertext{\item Optimization I:}
  X &= P_v^{*}\;\prod_{v \in W_v} P_v^{*} \ast P_w
  \intertext{\item Optimization II:}
  X &= \prod_{v,w \in W_v} P_v^{*}
  \end{align*}
\end{itemize}
where $\psi^t$ is the estimate at iteration $t$, $\phi_v$ is the observed data of view $v$, $P_v$ is the PSF of view $v$, and $P_v^*$ is the flipped PSF of view $v$.

\subsection* {2. Convergence and number of iterations}

The optimizations derived in \citet{Preibisch14} and listed above reduce the number of iterations the algorithm requires to converge. Convergence behavior of the different optimization variants were extensively studied in \citet{Preibisch14} and apply likewise to our implementation. In practice, choosing the number of iterations is a trade-off between achieved quality and computation time. We therefore leave it to the user, who needs to make this decision based on the particular situation (e.g. if deconvolution is performed in real-time, a reduced number of iterations might be preferred for an increase in overall acquisition speed). To facilitate the decision, we provide a tool for interactively investigating different numbers of iterations on a single cross-section (see also the Fiji plugin manual).

\subsection* {3. CUDA workflow for plane-wise multi-view deconvolution}

Our plane-wise multi-view deconvolution implementation uses multiple CUDA streams to overlap GPU computations with data transfer, such that not only copies to and from the GPU, but also loading and saving data from and to hard-drive come without additional cost. The implemented workflow is outlined below. Here, all processing and CUDA calls are asynchronous, i.e.~non-blocking. Synchronization is achieved by calls to \textit{cudaStreamSynchronize()}.\\ 

\begin{algorithm}[h!]
  \SetAlgoLined
  nStreams = 3\;
  \For{$z\leftarrow 0$ \KwTo $nStreams$}{
    Initialize $streams[z]$\;
    Load plane $z$ from hard-drive into main memory\;
  }
  \For{$z\leftarrow 0$ \KwTo $nStreams$}{
    $stream = streams[z \mod nStreams]$\;
    \If{$z >= nStreams$}{
        $cudaStreamSynchronize(stream)$\;
        Save deconvolved plane $(z-nStreams)$ to hard-drive\;
        Load plane $z$ from hard-drive into main-memory\;
    }
    Copy plane $z$ from main memory to GPU\;
    \For{$it\leftarrow 0$ \KwTo $nIterations$}{
      Calculate Richardson-Lucy step on $stream$\;
    }
    Copy plane $z$ from GPU to main memory\;
  }
  \For{$z\leftarrow (nPlanes - nStreams + 1)$ \KwTo $nPlanes$}{
    $stream = streams[z \mod nStreams]$\;
    $cudaStreamSynchronize(stream)$\;
    Save deconvolved plane $z$ to hard-drive\;
  }
  \caption{Workflow to interleave disk I/O and memory transfers with data processing.}
\end{algorithm}

\subsection* {4. Libraries and dependencies}
To efficiently calculate the Richardson-Lucy iteration step, convolutions were replaced by multiplications in Fourier domain. Fourier transformations were computed using the cuFFT library (https://developer. nvidia.com/cuFFT). Other arithmetic operations were implemented as custom CUDA kernel functions.

The entire workflow was implemented in the C programming language, using the CUDA specific extensions. The Fiji plugin was implemented in the Java programming language (Oracle Corporation). The C program was interfaced from Java using JNI (Java Native Interface).

The deployed plugin contains for each platform the corresponding binary library, which is statically linked agains the CUDA SDK. Additionally, the cuFFT library is bundled, which is required as a shared library.

Requirements for execution are a Nvidia graphics card that supports CUDA.

\pagebreak

\section*{Supplementary Table 1: Execution speeds using different graphics cards.}

\renewcommand{\tablename}{Supplementary Table}
\begin{table}[h!]
\begin{center}
\begin{tabular}  { lccc }    \toprule
                    & \multicolumn{3}{c}{Execution time (s)} \\ \cmidrule{2-4}
Graphics card       &  $512^3$ pixel  &  $1024^3$ pixel  &  $2048^3$ pixel  \\ \midrule
Quadro K2000        &           12.0  &            83.7  &           683.8  \\
Tesla C2075         &            6.6  &            48.2  &           378.7  \\
GeForce GTX 680     &            7.8  &            29.8  &           238.5  \\
Tesla K40c          &            3.9  &            19.4  &           153.6  \\
GeForce Titan black &            4.0  &            21.1  &           152.3  \\ \bottomrule
\end{tabular}
\end{center}
\caption{
Comparison of execution speed for plane-wise deconvolution using different graphics cards. 10 iterations have been performed deconvolving two views. Processing was performed on a Dell T6100 workstation (Intel E5-2630 @2.3 GHz 2 processors, 64 GB RAM). Data sizes correspond to the sizes padded for Fourier convolution, i.e. the sum of the actual data size and the size of PSF.
}
\end{table}
\FloatBarrier

\section*{Supplementary Figure 1: Deconvolution results viewed along the detection axis}
\renewcommand{\figurename}{Supplementary Figure}
\begin{figure}[h!]
\begin{center}
\includegraphics{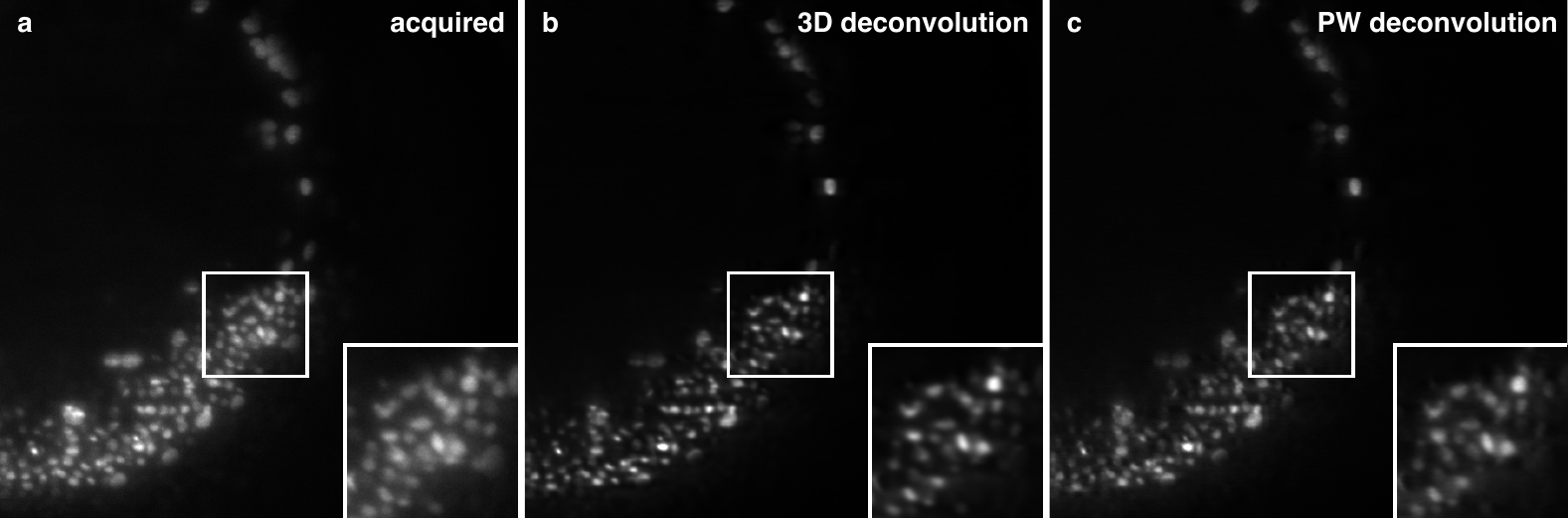}
\end{center}
\label{sfig:01}
\caption{
Deconvolution results viewed along the detection axis. \textbf{(a)} Acquired data of the first view of a 9 hours post fertilization old \textit{Tg(h2afva:h2afva-mCherry)} zebrafish embryo. \textbf{(b,c)} Multi-view deconvolution, performed \textbf{(b)} in full 3D and \textbf{(c)} plane-wise.
}
\end{figure}
\clearpage

\section*{Supplementary Figure 2: Comparison of various fusion methods using a simulated data set}
\renewcommand{\figurename}{Supplementary Figure}
\begin{figure}[h!]
\begin{center}
\includegraphics{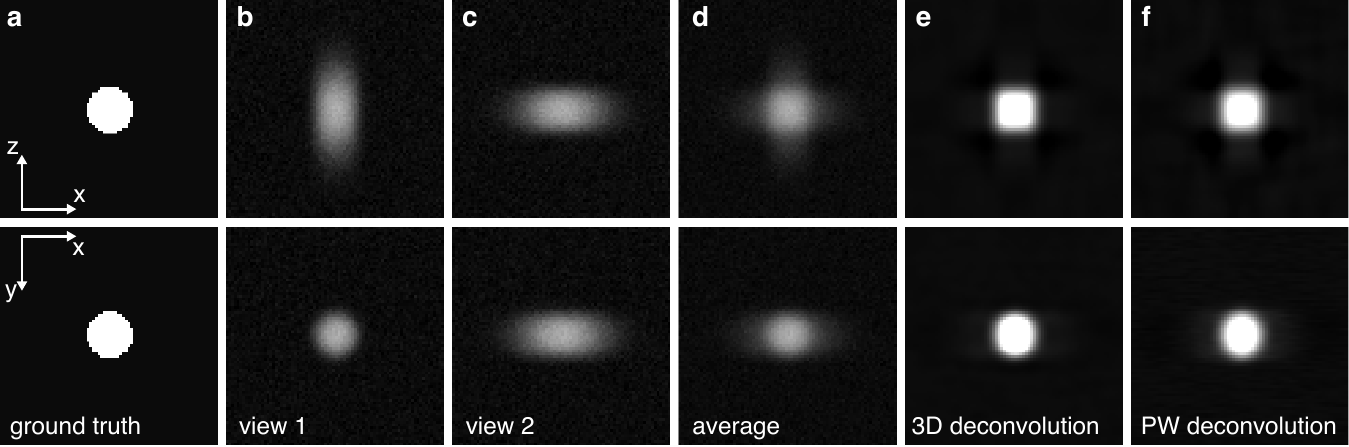}
\end{center}
\caption{Comparison of various fusion methods using a simulated data set. \textbf{(a)} Original data resembling a single nucleus of a \textit{Tg(h2afva:h2afva-mCherry)} zebrafish embryo. \textbf{(b, c)} Simulated view 1 and view 2, generated by convolving the original data with an elongated PSF in $z$- and $x$-direction. Fusion by \textbf{(d)} averaging, \textbf{(e)} 3D multi-view deconvolution and \textbf{(f)} plane-wise multi-view deconvolution.}
\label{sfig:02}
\end{figure}

\clearpage

\section*{Supplementary Figure 3: Comparison of deconvolution results under various noise levels}
\renewcommand{\figurename}{Supplementary Figure}
\begin{figure}[h!]
\begin{center}
\includegraphics{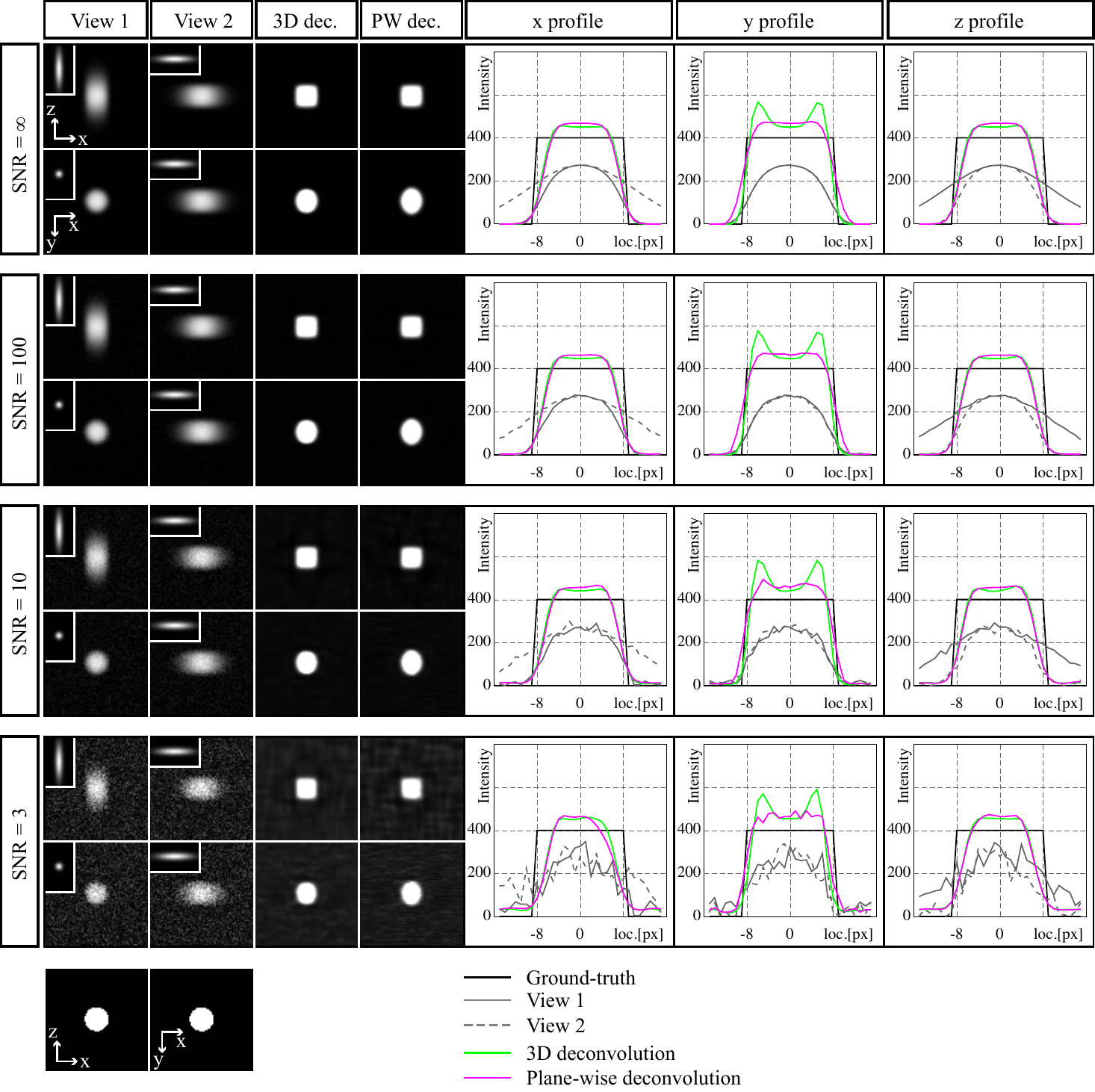}
\end{center}
\caption{
Comparison of deconvolution results under various noise levels. Different amounts of Gaussian noise were added to the simulated data from Supplementary Fig.~2. For each signal-to-noise ratio (SNR), both views and the deconvolution results from the 3D deconvolution and our plane-wise implementation are shown, along the rotation axis (top row) and along the detection axis of view 1 (bottom row). For each SNR value, line profiles are shown of the ground truth, the simulated data and the deconvolution results in all three dimensions. Throughout all tested SNR values, results of plane-wise deconvolution closely resemble the results of the original 3D implementation.
}
\label{sfig:03}
\end{figure}

\clearpage

\section*{Supplementary Figure 4: Comparison of deconvolution results using different PSFs}
\renewcommand{\figurename}{Supplementary Figure}
\begin{figure}[h!]
\begin{center}
\includegraphics{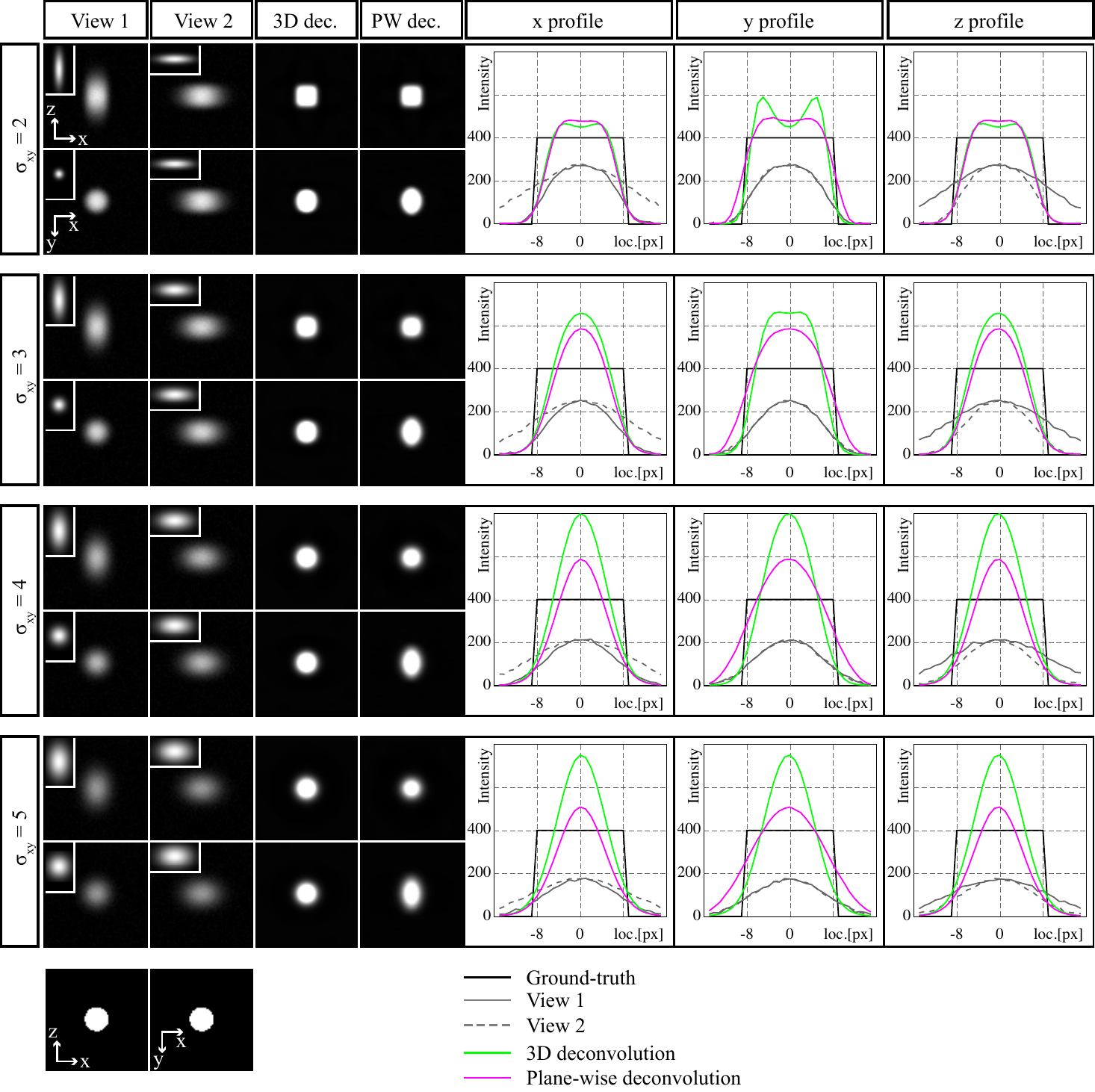}
\end{center}
\caption{
Comparison of deconvolution results using different PSFs. Simulated data were created as in Supplementary Fig. 2, using Gaussian PSFs with a fixed axial standard deviation $\sigma_{z}$ of eight pixels, as determined empirically on our microscope. Different values were used for the lateral standard deviation $\sigma_{xy}$. For each value, both views and the deconvolution results from the 3D deconvolution and our plane-wise implementation are shown, along the rotation axis (top row) and along the detection axis of view 1 (bottom row). Line profiles are shown of the ground truth, the simulated data and the deconvolution results in all three dimensions. While the results obtained by plane-wise and original 3D deconvolution are similar for small values of $\sigma_{xy}$ below a value of two, they start to diverge for higher values. $\sigma_{xy}$ on our microscopes was typically between 1.5 and 1.8 pixels.
}
\label{sfig:04}
\end{figure}
\FloatBarrier

\bibliographystyle{natbib}

\end{document}